\documentclass[a4paper,12pt]{article}
\usepackage{a4wide}
\usepackage[latin2]{inputenc}
\usepackage[dvips]{graphicx}
\usepackage[mathcal]{eucal}
\usepackage{amsfonts}
\usepackage{amsmath}
\usepackage{amsthm}
\usepackage{booktabs} 

\title{Physical space description of decorated quasicrystals}

\author{ 
\textsc{Pawe\l{} Buczek} 
\footnote{Current address: Department of Physics, Avadh Bhatia Physics Laboratory, University of Alberta,
Edmonton T6G 2J1, Alberta, Canada (also address for correspondence).}
\footnote{Electronic address: \texttt{\small{pbuczek@phys.ualberta.ca}}} , 
\textsc{Janusz Wolny}
\footnote{Electronic address: \texttt{\small{wolny@novell.ftj.agh.edu.pl}}} \\
\textit{\small{Faculty of Physics and Nuclear Techniques, AGH-UST}} \\ 
\textit{\small{al. Mickiewicza 30, 30-059 Krak\'ow, Poland}} 
}

\date{\today}

\renewcommand{\vec}[1]
{
{\bf #1}
}

\begin{document}

\maketitle

\begin{abstract}

In this paper the systematic method of dealing with the arbitrary decorations of quasicrystals is presented. The method is founded on the average unit cell formalism and operates in the physical space only, where each decorating atom manifests itself just by an additional component of the displacement density function in the average unit cell. Such approach allows us to use almost all classical crystallography algorithms for structure refining based on experimental data and may meaningly decrease the number of parameters which have to be fit. Further help for such analysis may be the use of proposed recently average Patterson function, here applied to decorated sets. As an example we present a description of a class of decorated quasicrystals based on Sturmian sequence of two interatomic spacings: we calculate explicitly structure factor, the shape of average Patterson function and give an algorithm for pattern analysis.

\end{abstract}

\begin{sloppy}

\section{Introduction}
\label{sec:Introduction}

Discovery of quasicrystals opened a completely new world for crystallographers but in the same moment left them almost unarmed -- all sophisticated and powerful methods developed to deal with strictly periodic structures appeared to be limited and insufficient for the description of this new long range atomic orderings. Elegant mathematical constructions gave us deep insight into these structures, but they scarcely very handy when one has to deal with real data obtained from scattering experiments. We focus here on problem of decorations. Obviously there is no theoretical difficulty to analyze arbitrary decoration using standard projection techniques, nevertheless the most common usage of atomic surfaces has to deal with their extremely complicated shapes and numerous fittable parameters. 

In this paper we propose a technique, which by operating solely in the physical space opens the possibility of reducing the number of parameters which have to be fit. Our method originates from the concept of reference lattice suggested in \cite{Wolny1998PhM}. Each of the decorating atoms appears as an additional component in so called average unit cell, which has very similar interpretation to that of the unit cell for periodic crystals; this allows for the utilization of the classical crystallography refining techniques in case of quasicrystals. As a useful tool we propose generalization of average Patterson function approach (proposed recently in \cite{Wolny2004}) to the case of decorated structures. Description of quasicrystals using multidimensional Patterson analysis has been already developed in \cite{Steurer1999b}, but our approach is rooted in different formalism. Our work may be understood as generalization of \cite{Kozakowski2003}.

The paper is organized as follows. First a short survey through the average unit cell formalism is given, which is subsequently used for systematic description of decorations. Using the new tool we give an algorithm of a structure refining for a class of 1D quasicrystals founded on Sturmian sequence of two interatomic spacings. The work ends with specific example.
\section{The formalism}
\label{sec:TheFormalism}

The concept of the reference lattice has been already proposed in \cite{Wolny1998PhM}, here we present only only the key ideas and features which we are going to utilize in the paper.

Suppose that $\Lambda$ is an orthogonal projection of Delon\'e set $\Lambda' \subset \mathbb{R}^{n}$, where $n$
is arbitrary, onto line $l$:
\begin{align}
\Lambda = \Pi_{l} (\Lambda'),
\end{align}
so $x_{n} \in \Lambda$ is a sequence of collinear points lying on $l$. $\Lambda$ may constitute Delon\'e set itself or even periodic crystal.

If there exist $n$-dimensional Fourier transform of set $\Lambda'$ there exists also 1D Fourier transform of a set $\Lambda$, being essentially a cut through a $k$-space of $\Lambda'$ along direction $l$. The intensity measured along this direction is:
\begin{align}
I(k)=\lim_{N \to \infty} \frac{1}{N^2} \Big| \sum_{n=1}^{N} f_{n} \exp(ikx_{n}) \Big|^{2},
\end{align}
where $f_{n}$ is a scattering power of the given atom in $\Lambda'$.

\subsection{Structure factor}
\label{subsec:StructureFactor}

In the average unit cell formalism for one type of atom in the system (with scattering power $f$) normalized intensity can be written as
\begin{align}
I(k) = | F(k) |^{2},
\end{align}
where the structure factor
\begin{align}
F(k) &=f \lim_{N \to \infty} \frac{1}{N} \sum_{n=1}^{N} \exp (i k x_{n}) 
=f \lim_{N \to \infty} \frac{1}{N} \sum_{n=1}^{N} \exp (i k u_{n}) 
=f \int_{- \infty}^{\infty} P_{\lambda} (u) \exp (i k u) \textrm{d} u.
\label{eq:1pFac}
\end{align}
We call the series $u_n$ the \textit{displacements sequence} of $x_n$ (induced by the reference lattice with period $\lambda = 2 \pi / k$):
\begin{align}
u_n=x_n - \left\| \frac{x_n}{\lambda} \right\| \lambda,
\end{align}
where $\left\|\cdot\right\|$ is the nearest integer
function: if $m \in \mathbb{Z}$ and $m \leq x < m+1$ then:
\begin{align}
\left\|x\right\|=
\left\{
\begin{array}{ll}
m & \textrm{if} \quad x \in [m,m+1/2), \\
m+1 & \textrm{if} \quad x \in [m+1/2,m+1).
\end{array}\right.
\label{eq:nif}
\end{align}

Any series $u'_n$ such that
\begin{equation}
u_n = u'_n - \left\| \frac{u'_n}{\lambda} \right\| \lambda,
\end{equation}
will be called an \textit{unreduced displacements sequence} (of $x_{n}$) if its elements belong to any finite interval.

Statistical distribution $P_{\lambda} (u)$ of distances $u_{n}$ will be called called \textit{displacement density function} (DDF). All considered DDFs are normalized to 1 and have value 0 outside interval $\left[-\lambda/2,\lambda/2\right]$. Statistical distribution $\widetilde{P}_{\lambda}(u)$ of $u'_{n}$ is called \textit{extended displacement density function} (EDDF) for $P_{\lambda} (u)$; we have:
\begin{align}
P_{\lambda}(u)=\sum_{m \in \mathbb{Z}} \widetilde{P}_{\lambda}(u+m\lambda).
\end{align}
$\widetilde{P}_{\lambda}(u)$ function may be supported on the finite interval different from $\left[-\lambda/2,\lambda/2\right]$, but it is still normalized. Note that we can use $\widetilde{P}_{\lambda}(u)$ instead of corresponding $P_{\lambda} (u)$ in all Fourier integrals like (\ref{eq:1pFac}). In general for each $k$ we should construct different $P_{\lambda}(u)$.

There is a handy theorem we will utilize in future, it comes from \cite{Buczek2005a}.
\newtheorem{theo}{Theorem}
\begin{theo}
\label{theo:sumOfDis}
Let $x_{n}=\alpha_n+\beta_n$ be a sum of two real series. If $d_{n}$ is a displacements sequence of $\alpha _{n}$
induced by a reference lattice, then
\begin{align}
u'_{n} &= d_{n} + \beta _{n} \label{eq:sumOfDis}
\end{align}
is an unreduced displacements sequence of $x_{n}$ induced by the same lattice.

\begin{proof}
Let $\lambda$ be the period of the reference lattice. We have to show that
\begin{align}
u_{n} = x_{n} - \left\| \frac{r_{n}}{\lambda} \right\| \lambda = u'_n - \left\| \frac{u'_n}{\lambda} \right\|
\lambda.
\label{eq:toPr1}
\end{align}

Note that for any real number $x,y$, $\left \| x - \left\| y \: \right\| \right \| = \left\| x \right\| - \left\| y \right\|$, since $\left\| y \right\|$ is an integer. We can write the right hand side of (\ref{eq:toPr1}) as
\begin{equation}
\beta_{n} + \alpha_{n} - \left\| \frac{\alpha _{n}}{\lambda}
\right\| \lambda - \left\| \frac{\beta_{n} + \alpha_{n} - \left\| \alpha _{n} /
\lambda \right\| \lambda}{\lambda} \right\| \lambda 
= x_{n} - \left\| \frac{x_{n}}{\lambda} \right\| \lambda,
\end{equation}
which is the left hand side.
\end{proof}
\end{theo}

The structure factor for $mk, \: m \in \mathbb{Z}$ can be computed from the reference lattice for $k$; it is the $m$-th Fourier mode of the distribution $P_{\lambda}(u)$. Thus, a single average unit cell is sufficient to analyze structures whose scattering occurs at multiples of a fixed scattering vector $k_0$. This situation includes, but is not limited to, the case where the original point pattern was periodic with period $2 \pi /k_0$.

For modulated structures (including quasicrystals), there are usually two periods, $a_{0}$ and $b_{0}$, which may be incommensurate. Using two reference lattices, the first one having periodicity $a_{0}$ and the second having periodicity $b_{0}$, the structure factor for the sum of two scattering vectors $k_{0} \equiv 2 \pi/ a_{0}$ and $q_{0} \equiv 2 \pi/ b_{0}$ can be expressed by:
\begin{align}
F(k_{0} + q_{0}) &= \lim_{N \to \infty} \frac{1}{N} \sum_{n=1}^{N} \exp(i(k_{0} + q_{0})x_{n}) 
= \lim_{N \to \infty} \frac{1}{N} \sum _{n=1} ^{N} \exp(i(k_{0} u_{n} + q_{0} v_{n})) \nonumber \\
&= \int _{-a_{0}/2} ^{a_{0}/2} \int _{-b_{0}/2} ^{b_{0}/2} P_{a_{0},b_{0}}(u,v) \exp(i(k_{0} u + q_{0} v)) \textrm{d}v \textrm{d}u, 
\label{eq:2pFac}
\end{align}
where $u$ and $v$ are the shortest distances of the atomic position from the appropriate points of two reference lattices and $P_{a_{0},b_{0}}(u,v)$ is the corresponding probability distribution, which thus describes a {\em two dimensional} average unit cell. Likewise, the structure factor for a linear combination $nk_{0} + mq_{0}, \:n,\:m \in \mathbb{Z}$ is given by the $(n,m)$ Fourier mode of $P_{a_{0},b_{0}}(u,v)$. This means that the average unit cell,
calculated for the wave vectors of the main structure and its modulation, can be used to calculate the peak intensities of any of the main reflections and its satellites of arbitrary order. Using (\ref{eq:2pFac}) and its generalization, it is possible to calculate the intensities of all peaks observed in the diffraction patterns.

\subsection{Patterson analysis}
\label{sebsec:PattersonAnalysis}

Phase inversion problem pushed to development of a variation of the above method called \textit{average Patterson function analysis}.

It is well known that intensity $I(k)$ can be obtained as a direct Fourier transform of the autocorrelation function
\begin{align}
G(x) &= \int_{-\infty}^{\infty} \rho(\xi) \rho(\xi + x) \mathrm{d} \xi
	= \int_{-\infty}^{\infty} \Big(\sum_{j} f_{j} \delta(\xi - x_{j})\Big)
		\Big( \sum_{k} f_{k} \delta (\xi + x - x_{k}) \Big) \mathrm{d} \xi \nonumber \\
	&= \sum_{j,k} f_{j} f_{k} \delta \big(x - (x_{k} - x_{j})\big).
\label{eq:G}
\end{align}

Let's suppose for a while that we deal with only one type of the atom ($f_{j}=f=$const) in the structure. To calculate a Fourier transform of $G(x)$ for scattering vector $k$ we can utilize average unit cell formalism as well: again by introduction of the reference lattice with periodicity $\lambda=2\pi/k$. Average Patterson function of the structure $\mathcal{G}_{\lambda}(u)$ is a distribution of values of all interatomic distances reduced to average unit cell. We may however first convey $x_{k}$ and $x_{j}$ separately to the unit cell and subsequently compute the distribution -- we may always add/subtract the multiplicity of the reference lattice to the distance; therefore $\mathcal{G_{\lambda}}(u)$ may be regarded as the distribution of distances between these points' images in the cell. Since displacements sequences of $x_{k}$ and $x_{j}$ can be treated as independent variables (the last sum in (\ref{eq:G}) is over all possible pairs of atoms) $\mathcal{G}_{\lambda}$ is essentially an autocorrelation of probability distribution $P_{\lambda}(u)$:
\begin{align}
\mathcal{G}_{\lambda}(x) = \int_{-\infty}^{\infty} P_{\lambda} (\xi) P_{\lambda} (\xi - x) \mathrm{d} \xi.
\label{eq:G1}
\end{align}
$\mathcal{G}_{\lambda}$ has all properties of $P_{\lambda}$; in particular we can use unreduced average Patterson functions where applicable and construct many-parameters average Patterson unit cells.

The intensity:
\begin{align}
I(k) = f^{2} \int_{-\infty}^{\infty} \mathcal{G}_{\lambda}(x) \exp(ikx) \mathrm{d} x.
\end{align}

\subsection{Reference sites}
\label{subsec:ReferenceSites}

Let us now consider decorated structures. We will adopt two very reasonable assumptions about atomic structure which can be observed in nature. First, every relevant physical system consists of finite number of types of atoms and we will denote this number by $N$. Second, the atoms tend to organize themselves on scales comparable to interatomic distances while in the large scales every atom should be seen as distributed uniformly, close to its average concentration (we do not consider such effects like precipitations). Note that these assumptions allow us to work (for example) with every decorated quasicrystal obtained from projection method and also with classical crystals.

We introduce idea of \textit{decoration sites}, $x_{n} \in \Lambda_{0}$, where $\Lambda_{0}$ fulfills assumption made upon $\Lambda$ in the previous section. Each decoration site has a given type, denoted by the first capital letters of alphabet ($A$, $B$, \ldots). Relative numbers of given types of sites are $\varphi_{X}$, where $X$ stands for $A$, $B$,\ldots $\varphi$'s are normalized:
\begin{align}
\sum_{X} \varphi_{X}=1.
\end{align}
To every type of site we ascribe a group of atoms, with well defined scattering powers and positions in relation to the site; number of atoms in such group is $n_{X}$ (again $X$ stands for $A$, $B$,\ldots) and
\begin{align}
\sum_{X} n_{X}=N,
\end{align}
what means that atoms with the same scattering powers are considered to be different type if they are ascribed to sites with a different type. We will refer to the given atoms using notation $X_{j}$ ($j=1,\ldots,n_{X}$), what reads: $j^{\mathrm{th}}$ atom in the group corresponding to site $X$. Scattering powers of atoms are $f_{X_{j}}$ and their relative positions with respect to the site $a_{X_{j}}$.

Classical crystallography uses only one type of sites organized in one of the Bravais lattice -- the above idea is a generalization of classical picture.

\subsection{Overall displacements density function}
\label{subsec:OverallDispalcementsDensityFunction}

Displacement density function for $X$-type sites is $\mathcal{P}_{\lambda}^{X}(u)$. Possibly  extended DDF for the atom of type
$X_{j}$ is simply

\begin{align}
\widetilde{P}_{\lambda}^{X_{j}} (u) = \mathcal{P}_{\lambda}^{X}(u - a_{X_{j}}),
\label{eq:partialDDF}
\end{align}
what follows from Theorem \ref{theo:sumOfDis}, where we put $\beta_{n} = a_{X_{j}} = \mathrm{const}$. We will refer sometimes to DDFs as ``images''. As a result we get that the cell image of the given type of the decorating atom is just shifted image of the given type of reference site. The main advantage of working exclusively in the physical space can be easily seen.


Now we are able build overall (E)DDFs $\widetilde{P}_{\lambda}(u)$, which Fourier transform given by (\ref{eq:1pFac}) will give us the structure factor for the whole decorated structure. In order to achieve that we must add all $\widetilde{P}_{\lambda}^{X_{j}}(u)$ weighted with scattering powers and relative number of given type of atoms:
\begin{align}
\widetilde{P}_{\lambda} (u) = \Big(\sum_{X} \sum_{j=1}^{n_{X}} f_{X_{j}} \varphi_{X} \mathcal{P}_{\lambda}^{X}
	(u - a_{X_{j}})\Big) / \mathfrak{N},
\label{eq:overallDDF}
\end{align}
where the normalization $\mathfrak{N}$ is
\begin{align}
\mathfrak{N} = \sum_{X} \sum_{j=1}^{n_{X}} f_{X_{j}} \varphi_{X}.
\label{eq:normalization}
\end{align}

Reduced $\widetilde{P}_{\lambda}(u)$ constitutes average unit cell for the whole decorated structure and
structure factor for scattering vector $k=2\pi m/\lambda$ is then:
\begin{align}
F(k) &= \mathfrak{N} \int_{-\infty}^{\infty}
\widetilde{P}_{\lambda}(u) \exp(iku) \mathrm{d}u.
\end{align}

\subsection{Decorations in the language of Patterson function}
\label{subsec:DecorationsInTheLanguageOfPattersonFunction}

Usage of average Patterson function is crucial during the structure refining based on experimental data due to the fundamental problem of phase lost.

According to (\ref{eq:G1}) $\widetilde{\mathcal{G}}_{\lambda}$ in an autocorrelation of $\widetilde{P}_{\lambda}$. Written explicitly:
\begin{align}
\widetilde{\mathcal{G}}_{\lambda}(x) = \frac{1}{\mathfrak{N}^{2}}\sum_{X} \sum_{Y} \sum_{j=1}^{n_{X}} \sum_{k=1}^{n_{Y}}
	f_{X_{j}}f_{Y_{k}}\varphi_{X}\varphi_{Y}
	\int_{-\infty}^{\infty} \widetilde{P}_{\lambda}^{X_{j}} (\xi) \widetilde{P}_{\lambda}^{Y_{k}} (\xi - x)
		\mathrm{d} \xi,
\label{eq:GExplicitly}
\end{align}
where $X$ and $Y$ run independently through $A$, $B$,\ldots We can see that in the average cell $N^2$ components $P_{\lambda}^{X_{j}}
\ast P_{\lambda}^{Y_{k}}$ may be pointed, each corresponding to a convolution of a given pair of decorating
atoms' images. Intensity is given by:
\begin{align}
I(k) = \mathfrak{N}^{2} \int_{-\infty}^{\infty} \widetilde{\mathcal{G}}_{\lambda}(x) \exp(ikx) \mathrm{d} x.
\end{align}

If $I(2 \pi m / \lambda)$, $n \in \mathbb{Z}$, are known, we can rebuild $\widetilde{\mathcal{G}}_{\lambda}(x)$ by means of Fourier series as follows:

\begin{align}
\widetilde{\mathcal{G}}_{\lambda}(x) = \frac{1}{\mathfrak{N}^2 \lambda} \sum_{m \in \mathbb{Z}} I(2 \pi m / \lambda) 
	\exp(-i2 \pi m x / \lambda).
\end{align}

Generalization to the case with many-parameters AUC is straightforward (compare to (\ref{eq:2pFac})); we give it here for 2 parameter unit cell:
\begin{align}
\widetilde{\mathcal{G}}_{a,b}(u,v) = \frac{1}{\mathfrak{N}^2 ab} \sum_{n_{1} \in \mathbb{Z}} \sum_{n_{2} \in \mathbb{Z}}
	I(2 \pi (n_{1} / a + n_{2} / b)) \exp(-i 2 \pi ( n_{1} u / a + n_{2} v / b)).
\label{eq:inverseTransform}
\end{align}
\section{Example: 1D modulated crystals}
\label{sec:Example}

In this section we are going to present the way how the formalism works for the class of 1D incommensurately modulated crystals. The method is of course not limited to 1D structures and the example was chosen for its clarity. Generalization to real (3D) structures is possible (as suggested in the very beginning): we always deal with effective 1D structures since structure factor contains scalar product of scattering vector and position vector, what corresponds to projection of the structure onto a given direction. Preliminary work shows that calculations for real 3D structures are characterized only by more complicated shapes of the distributions.

Let us consider an aperiodic Sturmian sequence of two (positive) distances $\mathcal{A}$ and $\mathcal{B}$ which sequence is coded using population vector $\vec{p}_{m}=(p_{m}^{\mathcal{A}},p_{m}^{\mathcal{B}})$, $p_{m}^{\mathcal{A}}+p_{m}^{\mathcal{B}}=m$ and $p_{m}^{\mathcal{X}}$ stands for number of $\mathcal{X}$-type elements in the first $m$ terms of the sequence:
\begin{align}
p_{m}^{\mathcal{A}} &= \Big\|\frac{m}{\sigma+1}\Big\|, \\
p_{m}^{\mathcal{B}} &= m - \Big\|\frac{m}{\sigma+1}\Big\|.
\end{align}
where $\sigma \in (0,1) \cap (\mathbb{R} \setminus \mathbb{Q})$. 

\subsection{Decoration sites}
\label{subsec:DecorationSites}

Set of decoration sites $\Lambda_{0}$ contains points of the form:
\begin{align}
x_{m}=\vec{p}_{m} \cdot (\mathcal{A},\mathcal{B}).
\end{align}

We will say that the decoration site is $X$-type if $x_{m+1}-x_{m}=\mathcal{X}$, i.e. if the next site is in the distance $\mathcal{X}$.

Since
\begin{align}
\lim_{m \to \infty}\frac{p_{m}^{\mathcal{A}}}{p_{m}^{\mathcal{B}}} = 1/\sigma,
\end{align}
the relative concentrations of the given type of sites are
\begin{align}
\varphi_{A} &= \frac{1}{1 + \sigma}, \quad \varphi_{B} = \frac{\sigma}{1 + \sigma}.
\end{align}

One can easily recognize that $\Lambda_{0}$ is just a model set obtained by projection a part of two dimensional square lattice onto a line with a slope $\tan^{-1}(\sigma)$.

We will take our basic distances to be equal to respectively 
\begin{align}
\mathcal{A} &=\frac{\kappa(1+\sigma)}{\kappa + \sigma}a_{0}, \quad \mathcal{B} =\mathcal{A}/\kappa,
\label{eq:AB}
\end{align}
$\kappa \in \mathbb{R_{+}\setminus} \{1\}$. $\sigma$ determines the sequence of $\mathcal{A}$ and $\mathcal{B}$ and $\kappa$ their relative length.

It can be shown that reference sites' positions are given by:
\begin{align}
x_{m}=ma_{0} + UM_{b_{0}}(ma_{0}),
\label{eq:refSites}
\end{align}
where:
\begin{align}
M_{b_{0}}(ma_{0})=\frac{1}{2}-\{x/b_{0}+\frac{1}{2}\}, \\ 
b_{0}=(\sigma+1)a_{0} \label{eq:sigma}, \quad U=\mathcal{A}-\mathcal{B},
\end{align}
and $\{x\} = x - \| x - \frac{1}{2} \|$ is the fractional part of $x$.

\subsection{Average unit cells for decoration sites}
\label{sec:AvUnitCell}

Form of eq. (\ref{eq:refSites}) suggests, that our structure is incomensurately modulated crystal with
main period $a_{0}$ and period of modulation $b_{0}$. For such a structure it has been shown (e.g. \cite{Yamamoto1982}), that diffraction
pattern is supported on a discrete set of $k$'s of a form:
\begin{align}
k=n_{1}k_{0} + n_{2}q_{0},
\label{eq:k_support}
\end{align}
where $k_{0}=2 \pi/a_{0}$ and $q_{0}=2 \pi/b_{0}$. To calculate the whole pattern we will have to construct 2-parameter
average unit cell $\mathcal{P}_{a_{0},b_{0}}(u,v)$, based on these scattering vectors.

Let's first build one parameters distributions. We can obtain that possibly unreduced
displacement sequences of the decorating sites positions for $k_{0}$ and $q_{0}$ are respectively (we utilize Theorem
\ref{theo:sumOfDis}):

\begin{align}
u'_{m} &= x_{m} - \Big\| \frac{x_{m}}{a_{0}} \Big\| = UM_{b_{0}}(ma_{0}), \\
v'_{m} &= x_{m} - \Big\| \frac{x_{m}}{b_{0}} \Big\| = \xi u'_{m},
\label{eq:correlation}
\end{align}
where :
\begin{align}
\xi=1-b_{0}/U=(1+\sigma)/(1-\kappa).
\label{eq:xi} 
\end{align}
Using Kronecker's theorem \cite{Hardy1962} we can state that both of these sequences are
uniformely distributed and hence corresponding (E)DDF's have just rectangular shape.

It is also easy to show, that given types of sites reduced to average unit cell always occupy its well defined parts, as shown on Figure \ref{fig:Figure2}. Parameters of theses distribution are $U$ and $V=\xi U$ for $k_{0}$ and $q_{0}$ respectively.
\begin{align}
U_{A} &= U \frac{1}{1+\sigma}, \quad U_{B} = U \frac{\sigma}{1+\sigma}; \\
V_{A} &= V \frac{1}{1+\sigma}, \quad V_{B} = V \frac{\sigma}{1+\sigma}
\end{align}

\begin{figure}[htb]
        \begin{center}
                \includegraphics[width=0.75\textwidth,angle=0]{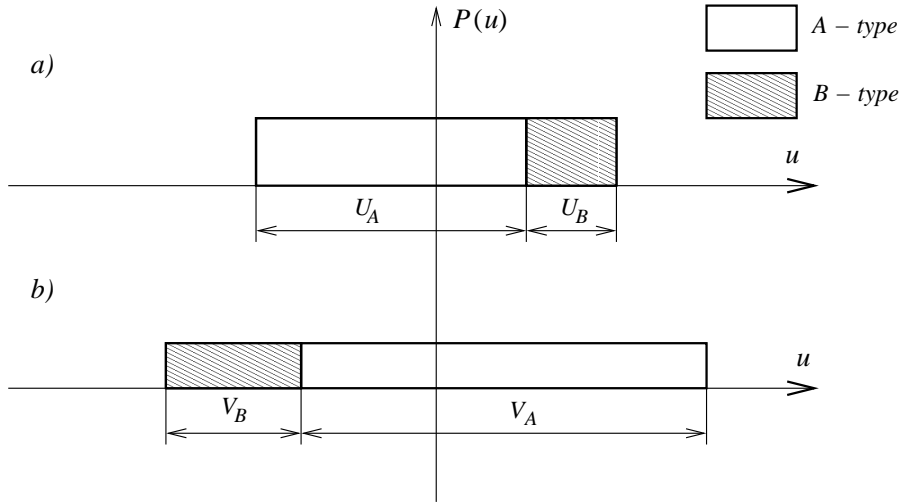}
        \end{center}
        \caption{Average unit cells for decoration sites. a) \& b) show (possibly unreduced) AUC's for $k_{0}$ and $q_{0}$ respectively. Given types of sites always occupy well defined parts of the cells. Picture shows situation for $\xi<0$. The distributions are centered around 0.}
        \label{fig:Figure2}
\end{figure}

\subsection{Overall density function}
\label{subsec:OverallDensityFunction}

Now let's consider 2-parameter average unit cell. Strong correlation between displacements sequences $u'_{m}$ and $v'_{m}$ given by (\ref{eq:correlation}) makes the 2-parameter distributions to be supported on fragments of straight lines only. Each decorating atom 
brings additional component to the overall function of the form (compare to (\ref{eq:partialDDF})):
\begin{align}
\widetilde{P}^{X_{l}}_{a_{0},b_{0}}(u,v) &= \frac{1}{U_{X}} \delta(v - \xi u - a_{X_{l}}(1-\xi)) 
R(u;U_{X},c_{X}+a_{X_{l}}). \\
c_{A} &=-\frac{U}{2}\frac{\sigma}{1+\sigma}, \quad
c_{B} =\frac{U}{2}\frac{1}{1+\sigma}
\end{align}
and $R(x;w,x_{0})=\theta(x-x_{0}+w/2)-\theta(x-x_{0}-w/2)$, where $\theta(x)$ stands for Heaviside step function. $R$ is rectangle with center $x_{0}$, width $w$ and heigth $1$. Example for 1 atom decorating $A$-type sites is shown in the Figure \ref{fig:Figure3}.

\begin{figure}[htb]
        \begin{center}
                \includegraphics[width=0.75\textwidth,angle=0]{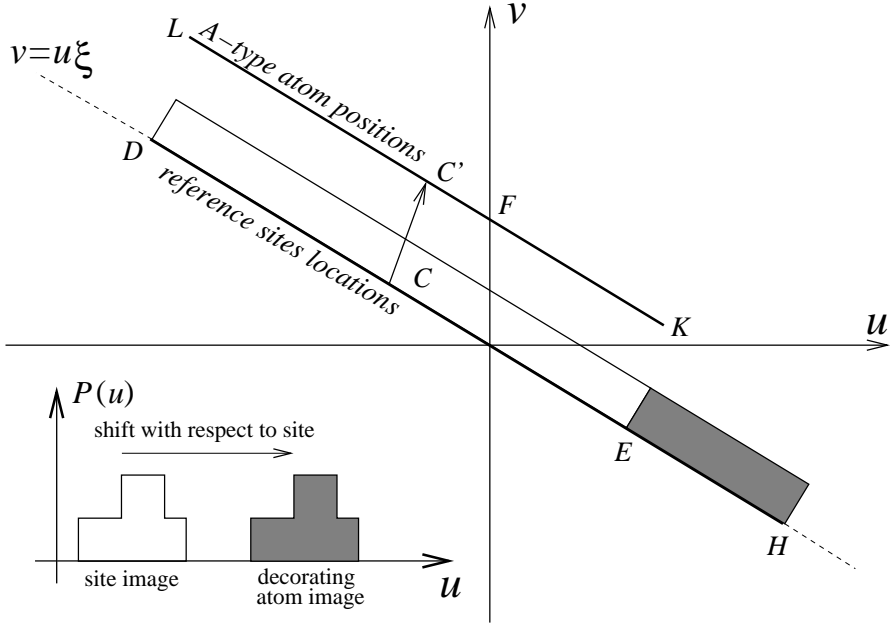}
        \end{center}
        \caption{Two parameter average unit cell for structure with 1 {$A$-type} atom decoration; also the positions of decoration sites were added (segments $DE$ and $EH$ are occupied by $A$ and $B$-type decoration sites respectively). Each decorating atom's image is an additional component $f_{X_{l}} \varphi_{X} \widetilde{P}^{X_{l}}_{a_{0},b_{0}}(u,v)$. In the discussed case segment $LK$ is an image of the decorating atom. $\overline{CC'}=(a_{A_{l}},a_{A_{l}})$ -- shift vector. $C=(c_{A},\xi c_{A})$, $D=(-U/2,-\xi U/2)$ and $F=(0,a_{A_{l}}(1-\xi))$. Such behavior is a consequence of image's shift in 1D unit cell (inner panel), please compare to equation (\ref{eq:partialDDF}).}
        \label{fig:Figure3}
\end{figure}

Once we have figured out geometry of the cell overall two parameter displacement density function can be written using (\ref{eq:overallDDF}):
\begin{align}
\widetilde{P}_{a_{0},b_{0}}(u,v) = \Big( \sum_{X} \sum_{l=1}^{x_{X}} f_{X_{l}} \varphi_{X} \widetilde{P}^{X_{l}}_{a_{0},b_{0}}(u,v) 
\Big) / \mathfrak{N},
\end{align}
where normalization $\mathfrak{N}$ is given by (\ref{eq:normalization}).

Structure factor for $k=n_{1}k_{0} + n_{2}q_{0}$ is given by:
\begin{align}
F(k)= \sum_{X} \sum_{l=1}^{n_{X}} f_{X_{l}} \varphi_{X}  F_{X_{l}},
\label{eq:finalFactor}
\end{align}
where:
\begin{align}
F_{X_{l}} &= \int_{-\infty}^{\infty} \int_{-\infty}^{\infty} \widetilde{P}^{X_{l}}_{a_{0},b_{0}}(u,v) \exp(i(n_{1}k_{0}u
+ n_{2}q_{0}v)) \mathrm{d}v\mathrm{d}u \nonumber \\ &=
-\frac{i}{KU_{X}}\mathrm{e}^{ika_{X_{l}}}(\exp(iK(c_{X}+U_{X}/2))-\exp(iK(c_{X}-U_{X}/2))) \nonumber \\ 
&= \exp(i(Kc_{X} + ka_{X_{l}})) \frac{\sin(w)}{w}.
\label{eq:FXl}
\end{align}

$K=n_{1}k_{0} + \xi n_{2}q_{0}=k-n_{2}q_{s}$, $q_{s}=q_{0}(1-\xi)$ and $w=KU_{X}/2$. Equation (\ref{eq:finalFactor}) is valid only for values of $k$ given by (\ref{eq:k_support}) and allows us to calculate intensity of each peak in the pattern. We can treat (\ref{eq:finalFactor}) also as continuous function of scattering vector $k$, parametrized by the order of the satellite $n_{2}$. In the latter case we obtain so called envelope functions $E_{m}(k)$, going through all satellites of the order $m=n_{2}$.

\subsection{Comment on factor $\mathrm{e}^{ika_{X_{l}}}$} 
\label{subsec:CommentOnFactor}

When we add an atom at position $x_{0}$ with respect to the decoration site, the corresponding part of DDF in 2-parameter AUC is will be just the shifted image of this decoration site; the shift vector reads $(x_{0},x_{0})$. Partial structure factor of this shifted DDF for $k=n_{1}k_{0} + n_{2}q_{0}$ is:
\begin{align}
F(k) &\sim \int_{-\infty}^{\infty} \int_{-\infty}^{\infty} \widetilde{P}(u-x_{0},v-x_{0}) \exp(i(n_{1}k_{0}u + n_{2}q_{0}v)) \mathrm{d}v\mathrm{d}u \nonumber \\
&= \int_{-\infty}^{\infty} \int_{-\infty}^{\infty} \widetilde{P}(u',v') \exp(i(n_{1}k_{0}(u'+x_{0}) + n_{2}q_{0}(v'+x_{0})) \mathrm{d}v'\mathrm{d}u' \nonumber \\
&= \mathrm{e}^{ikx_{0}} \int_{-\infty}^{\infty} \int_{-\infty}^{\infty} \widetilde{P}(u',v')
\exp(i(n_{1}k_{0}u' + n_{2}q_{0}v')) \mathrm{d}v'\mathrm{d}u'.
\end{align}
The last integral is of course proportional to structure factor for not shifted DDF. The relevance of the proportionality factor will be discussed  more deeply in the Summary.

\subsection{Patterson function}
\label{PattersonFunction}

As we have mentioned the Patterson function is of the fundamental importance. Generalizing equation (\ref{eq:GExplicitly}) to 2-parameter AUC we obtain:
\begin{align}
\widetilde{\mathcal{G}}_{a_{0},b_{0}}(u,v) = \frac{1}{\mathfrak{N}^{2}}\sum_{X} \sum_{Y} \sum_{j=1}^{n_{X}} \sum_{k=1}^{n_{Y}}
	f_{X_{j}}f_{Y_{k}}\varphi_{X}\varphi_{Y}
	(\widetilde{P}_{a_{0},b_{0}}^{X_{j}} * \widetilde{P}_{a_{0},b_{0}}^{Y_{k}}),
\end{align}
where the convolution $\widetilde{P}_{a_{0},b_{0}}^{X_{j}} * \widetilde{P}_{a_{0},b_{0}}^{Y_{k}}$ is given by:
\begin{align}
\frac{1}{U_{X}U_{Y}}&\int_{-\infty}^{\infty} \int_{-\infty}^{\infty} \delta(\beta - \xi \alpha - a_{X_{l}}(1-\xi))
	\delta(\beta - v - \xi (\alpha - u) - a_{Y_{k}}(1-\xi))	\times \nonumber \\ 
		&R(\alpha;U_{X},c_{X}+a_{X_{l}}) R(\alpha - u;U_{Y},c_{X}+a_{X_{l}}) \mathrm{d} \alpha \mathrm{d} \beta =\nonumber \\
		=&\delta(v - \xi u - (1 - \xi)(a_{X_{l}} - a_{Y_{k}})) T(u;l,w,h).
\label{eq:PattersonComponent}
\end{align}
$T(u;l,w,h) = \int_{-\infty}^{\infty} R(\alpha;U_{X},c_{X}+a_{X_{l}}) R(\alpha - u;U_{Y},c_{X}+a_{X_{l}}) \mathrm{d} \alpha$ is a trapezium-like function; in our case $l=||U_{X}|-|U_{Y}||$, $w=|U_{X}|+|U_{Y}|$, $h=2\min(|U_{X}|,|U_{Y}|)/(l+w)$. $T$ was depicted in the Figure \ref{fig:TFunction}.

\begin{figure}[htb]
        \begin{center}
                \includegraphics[width=0.75\textwidth,angle=0]{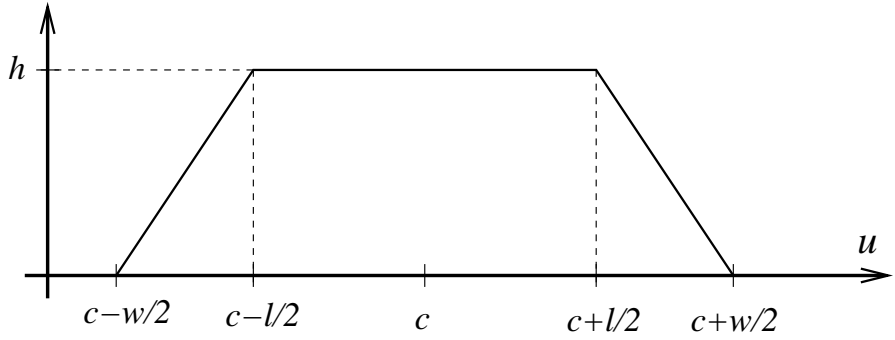}
        \end{center}
        \caption{Trapezium-like function used in equation (\ref{eq:PattersonComponent}). Average Patterson unit cell contains of sum of such functions with different parameters supported on segments of straight lines.}
        \label{fig:TFunction}
\end{figure}

Average Patterson unit cell contains $N^{2}$ components of the form given by (\ref{eq:PattersonComponent}); as ordinary average unit cell Patterson function function is supported on the fragments of straight lines only with slope $\xi$, but has more complicated shape.

\subsection{Pattern analysis}
\label{subsec:PatternAnalysis}

In this section we present the algorithm allowing recovery of the structure if the diffraction pattern $I(2 \pi (n_{1} / a_{0} + n_{2} / b_{0}))$ is known and we assume the crystal has internal structure as discussed above.

The pattern itself gives us values of $k_{0}$ and $q_{0}$, and hence $a_{0}$ and $b_{0}$; using (\ref{eq:sigma}) we can compute $\sigma$. Using (\ref{eq:inverseTransform}) we can inverse the transformation and obtain $\mathcal{G}_{a_{0},b_{0}}(u,v)$. Immediately a value of $\xi $ can be found, and subsequently using relation (\ref{eq:xi}) the ratio $\kappa = \mathcal{A}/\mathcal{B}$. Subsequently formulas (\ref{eq:AB}) give values of $\mathcal{A}$ and $\mathcal{B}$.

Average Patterson function can now be used to determine number of atoms, their types (scattering powers) and positions (type of decoration site and relative positions); of course due to phase problem we cannot avoid certain ambiguity. Detailed (numerical) algorithm will be discussed elsewhere. Many information can be extracted form average Patterson function relatively easily: note for example that from (\ref{eq:PattersonComponent}), e.g. by investigating points where Patterson function crosses axes $u$ or $v$, we can obtain information about relative positions of decorating atoms ($a_{X_{k}}-a_{Y_{l}}$).

Since average Patterson function is usually known with serious error, further refinement of atomic positions can be made by utilizing the structure factor. We can try direct fitting of the decorating distances including all of the diffraction peaks at once (e.g. by means of least squares method). It is also possible to use only subset of the pattern and optimize the fitness of envelopes $E_{m}(k)$.

As it was mentioned earlier structure factor can be described as a sum of components $F_{X_{l}}$ given by (\ref{eq:FXl}), which are products of two parts. The first one is the discussed phase factor, which depends only on relative decorating atom position; the second one is characteristic for the structure and does not depend on decoration. It is important feature which may improve the numerical algorithm. The part independent on decoration has to be calculated only once and it is enough to limit the refinement to the phase factors. 

\subsection{Specific example}
\label{SpecificExample}

As an example we present a structure with the following parameters: $\sigma= \pi$, $\kappa = \sqrt{2}$ and values of $\mathcal{A}$ and $\mathcal{B}$ are chosen such that $a_{0}=1+1/\tau^{2}$, $\tau$ is a golden ratio equals $(1 + \sqrt{5})/2$. The reason for such strange-looking choice is to make correspondence with other publications; conventionally for $\sigma=1 / \tau$ and $\kappa = \sigma$ we get famous Fibonacci chain. Above choice determines the value of other parameters, gathered in Table \ref{tab:parameters}.

\begin{table}[htbp]\label{tab:parameters}
	\centering
		\begin{tabular}{@{}ccc@{}} \toprule
			& \multicolumn{2}{c}{Value} \\ \cmidrule(rl{0.75em}){2-3}
			parameter & \multicolumn{1}{c}{exact} & \multicolumn{1}{c}{approximate} \\ \midrule
  		$\mathcal{A}$ & $\frac{\sqrt{2} (1 + \pi)}{\sqrt{2} + \pi} (1+1/\tau^{2})$ & 1.777  \\ 
  		$\mathcal{B}$ & $\mathcal{A} / \kappa$ & 1.256    \\ 
  		$b_{0}$ & $(1 + \pi) (1+1/\tau^{2})$ & 5.724    \\ 
  		$\xi$ & $\frac{1 + \pi}{1 - \sqrt{2}}$ & $-9.999$ \\ 
  		$k_{0}$ & $2 \pi / (1+1/\tau^{2})$ & 4.547    \\ 
		  $q_{0}$ & $k_{0} / (1 + \sigma)$ & $1.098$ \\ 
  		$q_{s}$ & $q_{0}(1-\xi)$ & $12.074$ \\ \bottomrule
		\end{tabular}
	\caption{The parameters of the system, which was used as the example.}
\end{table}

%

Decoration consists of two identical atoms with scattering powers equal to 1: the first occupies the the $A$-type segment in position $a_{A_{1}} = 0$, the second $B$-type segment in position $a_{B_{1}} = \mathcal{B}/2 \approx 0.628$. 

Average 2-parameter Patterson function was plotted in the Figure \ref{fig:exPatterson}. Please refer to the caption for detailed description. Figure \ref{fig:exPattern} presents diffraction pattern of the obtained structure with envelope functions.


\begin{figure}[htb]
        \begin{center}
                \includegraphics[width=0.75\textwidth,angle=0]{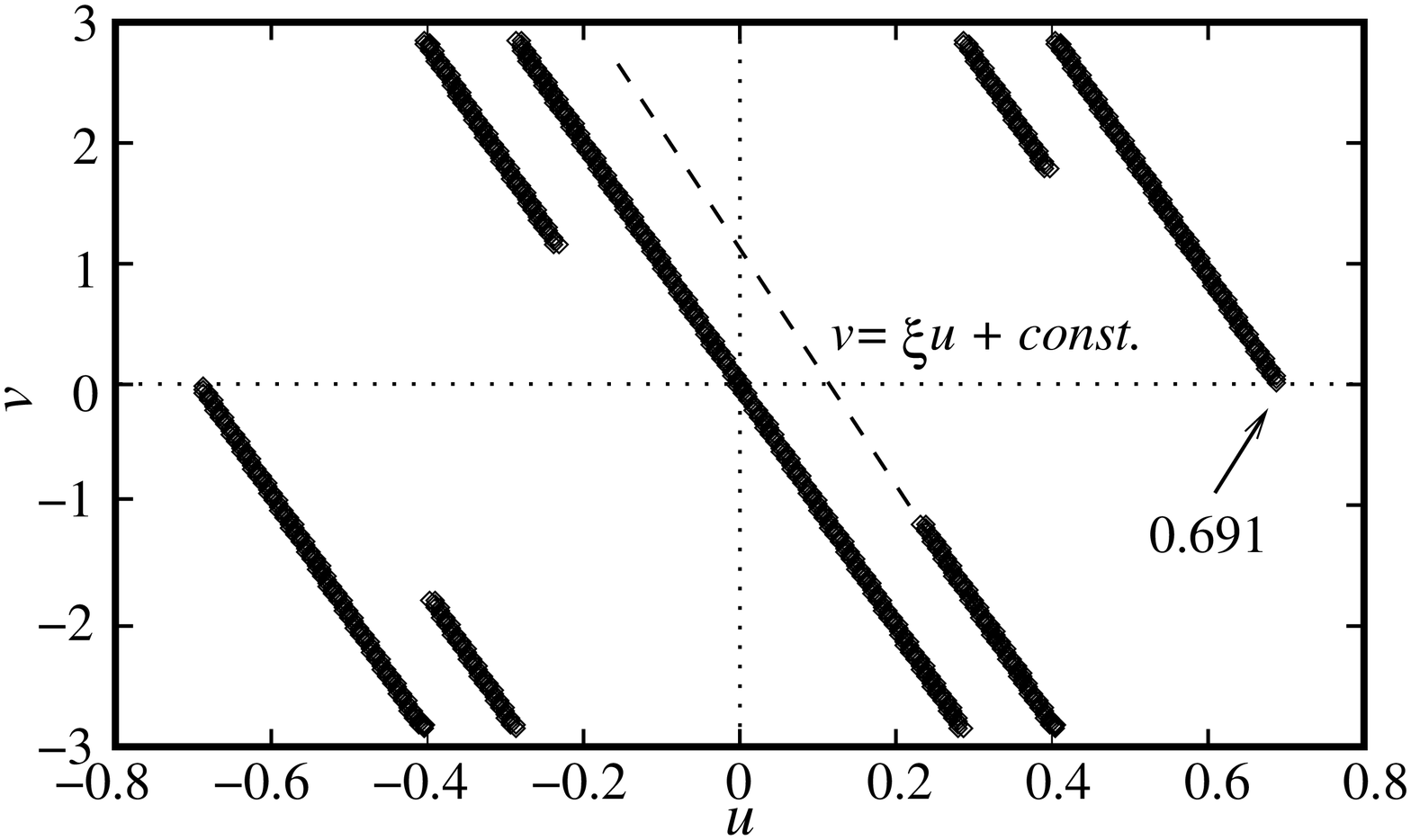} ~a)
                \includegraphics[width=0.75\textwidth,angle=0]{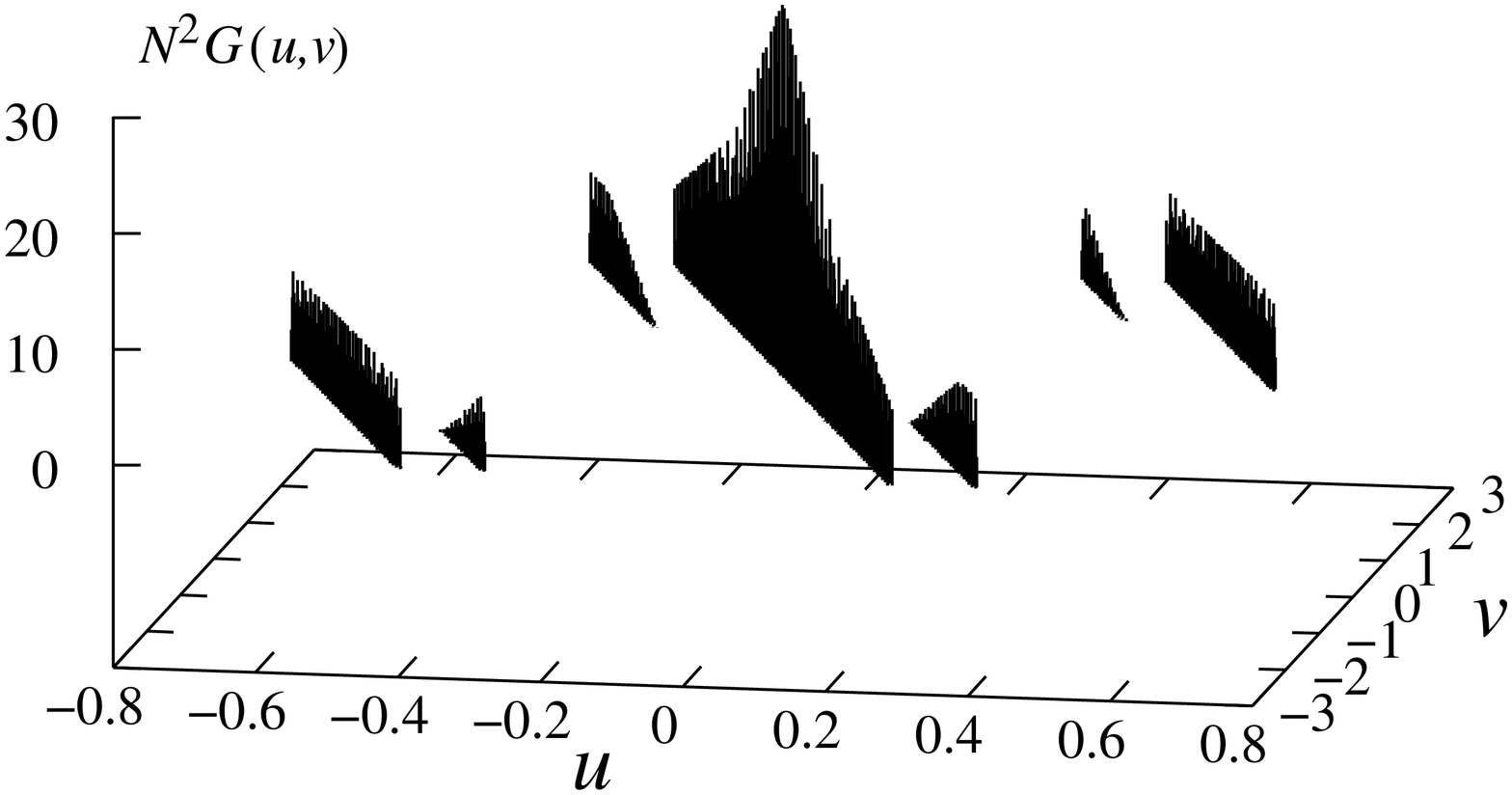} ~b)
        \end{center}
        \caption{Average Patterson function for the discussed example. Panel a) shows support of the function: it is a set of line segments with common slope $\xi$. Support crosses $u$ axis for $u=0$ and $u \approx \pm 0.691$, what corresponds to all possible values $(1 - \xi) (a_{A_{1}} - a_{B_{1}}) / \xi = \pm \frac{1}{2} a_{0}$ (in this case). Panel b) presents perspective view of the cell itself. Results are numerical (histogram for about 20000 atoms in the structure) and fully supports analytical calculations.}
        \label{fig:exPatterson}
\end{figure}

\begin{figure}[htb]
        \begin{center}
                \includegraphics[width=0.75\textwidth,angle=0]{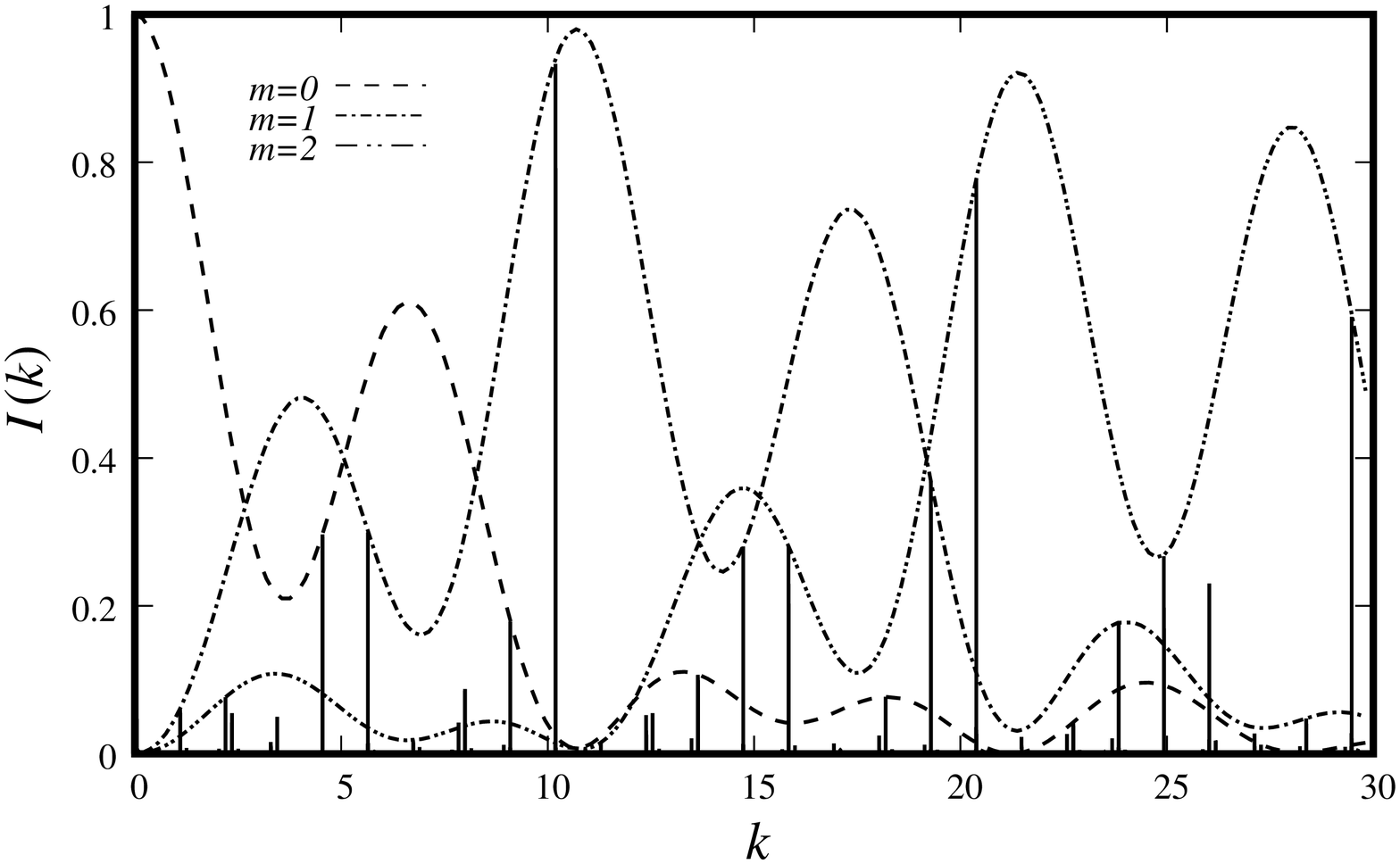}
        \end{center}
        \caption{Diffraction pattern for presented example. The envelope functions of orders $m=0$, 1 and 2 were plotted. Pattern comes from numerical experiment, while envelopes were calculated analytically; full agreement exists for all diffraction peaks with given $m$.}
        \label{fig:exPattern}
\end{figure}

\section{Summary}
\label{sec:Summary}

The statistical approach to quasicrystals' decoration which has been presented in this paper can be intuitively understood as extension of the classical terms to the realm of non-periodic orderings. Average unit cell is extension of the classical unit cell; decorating atoms are represented not as $\delta$-like objects in the cell, but correspond to non-discrete ``images''. Once we allow such generalization of standard notions, we can use the new concepts almost in the same manner as we used to work with the unit cell in the strictly periodic case.

Structure factor is constructed in the way similar to classical case, as a sum of Fourier transform of similar objects with adequate phase factors ($\exp(ika_{X_{l}})$, for the example discussed), corresponding to atoms' positions. Such explicit separation of the decoration dependence from general properties of the structure can be a help during the refinement process.


Simultaneous usage of average Patterson function and structure factor during analysis of the patterns can split the process into two separate parts. Patterson function gives us information about structure properties independent of the decoration. It can also provide us with number of decorating atoms and types of sites they are adherent to. Positions refinement can be done using structure factor (aided by the envelope functions).



\bibliographystyle{unsrt}
\bibliography{quasi_bibliography}

\end{sloppy}
\end{document}